\documentclass[preprint]{elsart}

\usepackage{graphicx}              

\newcommand{\loo}{\,\raisebox{-.5ex}{$\stackrel{<}{\scriptstyle\sim}$}\,}
\newcommand{\be}{\begin{equation}}
\newcommand{\ee}{\end{equation}}

\begin{document}

\begin{frontmatter}

\title{The decay time scale for highly excited nuclei as seen from 
asymmetrical emission of particles}

\author[cyctamu,iopsas]{M.~Jandel},  \thanks{\footnotesize E-mail address: 
jandel@comp.tamu.edu }
\author[cyctamu,INR]{A.S.~Botvina}$^{}$,
\author[cyctamu]{S. J. Yennello},
\author[cyctamu]{G. A. Souliotis},
\author[cyctamu]{D. V. Shetty},
\author[cyctamu]{E. Bell},
\author[cyctamu]{A. Keksis}

\address[cyctamu]{Cyclotron Institute,
           Texas A$\&$M University, College Station, TX 77843, USA }
\address[iopsas]{Institute of Physics, Slovak Academy of Sciences, 
           Dubravska 9, 84228 Bratislava, Slovakia }

\address[INR]{Institute for Nuclear Research, Russian Academy of Science, 
              117312 Moscow, Russia}

\begin{abstract}

A novel method was developed for the extraction of short emission 
times of light particles from the projectile-like fragments in peripheral 
deep-inelastic collisions in the Fermi energy domain. 
We have taken an advantage of the fact that in the external Coulomb field 
particles are evaporated asymmetrically. 
It was possible to determine the emission times in the interval 50-500 fm/c 
using the backward emission anisotropy of $\alpha$-particles relative to the
largest residue, in the reaction $^{28}$Si + $^{112}$Sn at 50 MeV/nucleon. 
The extracted times are consistent with predictions based on the 
evaporation decay widths calculated with the statistical evaporation model 
generalized for the case of the Coulomb interaction with the target. 

\end{abstract}


\end{frontmatter}

\section{Introduction}
Strong external fields may substantially change the nature of physical 
processes as compared to ones which occur in an isolated environment.  
Studies of electromagnetic processes in 
relativistic heavy-ion collisions (see e.g. \cite{Bertulani88,Pshenichnov98} 
and references therein), demonstrate the possibility of 
a large energy transfer to nuclei.
It is natural that similar electromagnetic processes 
may occur for heavy ion collisions near the Fermi energy. In the case of 
deep-inelastic peripheral nucleus-nucleus collisions, the electromagnetic 
interaction is small compared to the nuclear one, however, it may 
provide an important contribution to fragment production leading to new 
effects. For example, as was pointed out in refs. \cite{Botvina99,Botvina01}, 
the fast multifragmentation of projectile-like and target-like sources is 
influenced by the Coulomb interaction between them, and this interaction 
leads to asymmetrical emission of intermediate mass fragments from the 
sources. 

The interplay between electromagnetic and nuclear forces should be 
also manifested in other phenomena. 
The evaporation of isolated nuclei is well described by present theories 
\cite{Weisskopf,Ericson}. However, in the case of collisions of nuclei, 
one may expect that the de-excitation of the projectile takes 
place in the vicinity 
of the target. In the peripheral deep-inelastic collisions, when 
the projectile-like nucleus acquires excitation energies 
E*$\sim$3-5 MeV/nucleon, the 
fast evaporation takes place on a time scale of 10$^{-22}$ s (50-500 fm/c). 
In this paper, we consider the effect of asymmetrical light particle 
evaporation, which may appear as a consequence of the Coulomb interaction 
between the highly excited nuclei that decay rather quickly in the vicinity 
of each other. We demonstrate that this effect can be used for 
determination of the decay time of these nuclei, as was previously 
discussed in ref. \cite{Hudan03}.

 \section{Experimental observation}

We reanalyzed the data obtained in the peripheral reactions of 
$^{28}$Si with $^{112}$Sn at 50 MeV/nucleon \cite{Laforest99}. 
The excited projectile-like source (PLS) was carefully reconstructed 
on event-by-event basis. 
We considered events containing the largest residue charge $Z(LR) \geq 7$
and light charged particles, having 
the total charge of the reconstructed projectile-like source
$Z(PLS)=$13 and 14.
These products are in the kinematics region of 
the projectile, with the laboratory velocities around $v \approx 0.3c$
We concentrate on events having $\alpha$-particles, since these particles 
are usually produced in the evaporation process. 

The isotopic resolution in the experiment was achieved only for $Z \leq 5$
and no free neutrons were detected. 
The reconstruction of the projectile-like source excitation energy
is therefore influenced by the uncertainty in the determination of the mass
of projectile-like source. However, in this work, 
the mass assumption for $Z > 5$ fragments used in the experimental data was applied 
to the simulation. Additionally, 
neutrons are omitted when reconstructing the excitation energy so that the 
simulated and experimental values of the excitation energy match. In this 
manner, we defined the apparent excitation energy
\begin{equation}
E^{*}_{app}=\sum_{i}(T^{PLS}_{i}+\Delta M_{i})-\Delta M_{PLS},
\label{for1}
\end{equation}
where $T^{PLS}_{i}$ are the kinetic energies of the fragments in c.m.s. of the  
projectile-like source, $\Delta M_{i}$ are 
the mass excesses of fragments and $\Delta M_{PLS}$
is the mass excess of the projectile-like source. 
The angular distribution of $\alpha$-particles was
extracted from parallel and perpendicular velocities
in the reference frame of the heaviest fragment, 
(which in our case was limited to $Z \geq 7$ to assure projectile-like events).  

\begin{figure}

\resizebox{1.0\textwidth}{!}{
\includegraphics{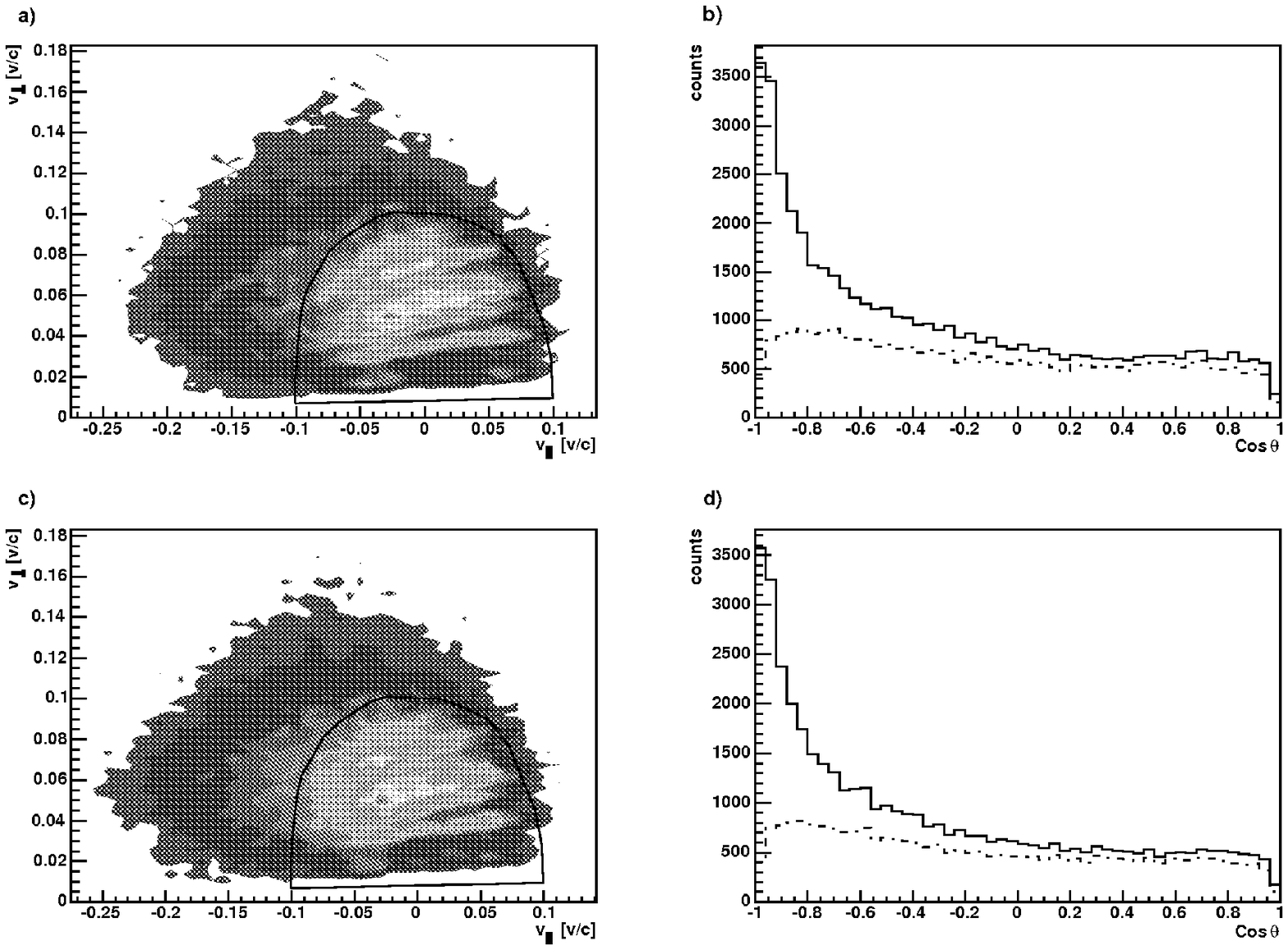}
}

\caption{Velocity scatter plots of $^{4}$He particles emitted from the PLS (a,c)
and angular distributions
of $\alpha$-particles (b,d) from the reaction $^{28}$Si + $^{112}$Sn at 50 MeV/nucleon. 
Angular distributions
shown by dotted lines correspond to $\alpha$-particles emitted 
within the Coulomb circle (shown by continuous line
in the scatter plots). 
Solid line represents all $\alpha$-particles from events
 that met the sum Z(PLS) criterion.
The results for 
Z(PLS)=14 and Z(PLS)=13 are shown on the top and bottom, respectively.}

\label{figure1}

\end{figure}

Fig. \ref{figure1} shows the scatter plots and angular distributions of 
emitted $\alpha$-particles relative to the largest fragment,
for
reconstructed projectile-like events with the charge $Z(PLS)=$14, 13 in the
reaction $^{28}$Si + $^{112}$Sn at 50 MeV per nucleon.
The asymmetrical emission is clearly observed.
We performed the additional selection of events inside the Coulomb circle
(shown by lines in the scatter plots in Fig. \ref{figure1}a, c),
in order to be more stringent and restrict our analysis
exclusively to the projectile-like evaporation process and avoid 
contributions from other processes.
The asymmetrical emission is still present as shown by 
dotted spectra in Fig. \ref{figure1}b, d. 

\begin{figure}

\resizebox{1.0\textwidth}{!}{
\includegraphics{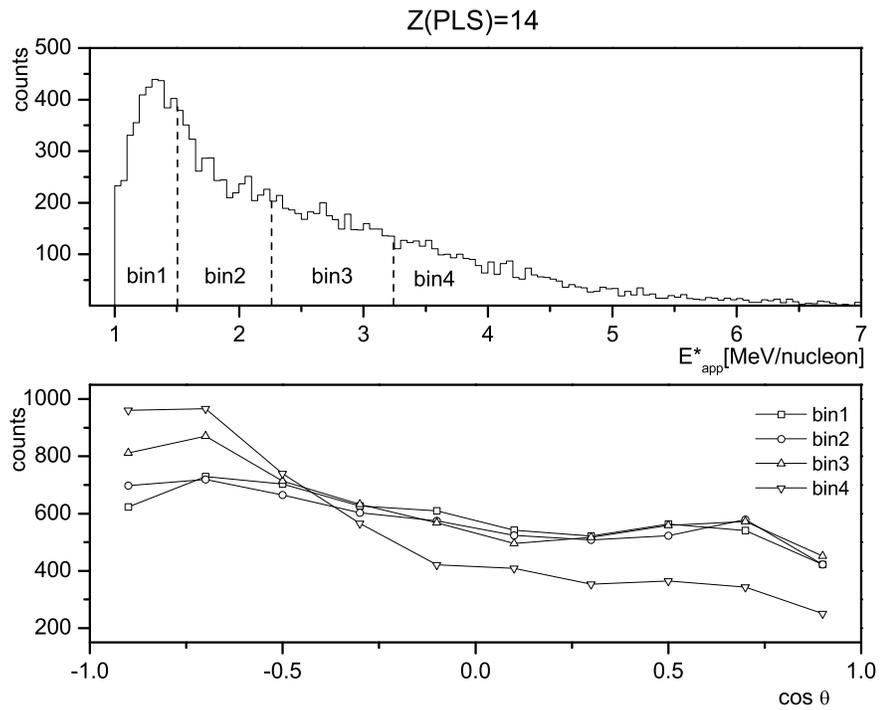}
}

\caption{The distribution of apparent excitation energy $E^{*}_{app}$ of reconstructed
projectile-like sources with the sum charge Z(PLS)=14 (upper) and angular distributions
of $\alpha$-particles in the c.m.s. of projectile-like fragment (lower) for selected $E^{*}_{app}$
bins.}
\label{figure2}
\end{figure}

Fig. \ref{figure2} shows the distribution of the PLS apparent 
excitation energy for the events in the Coulomb circle 
and how the angular asymmetry of emitted $\alpha$-particles 
changes with the excitation energy. 
The backward emission of $\alpha$-particles prevails over the central and 
forward emission, for 
all four bins of $E^{*}_{app}$ and more importantly,
the asymmetrical emission is increased with the excitation energy.

In order to estimate the angular asymmetry of $\alpha$-particles,
we defined three regions of emission: backward = cos $\theta \in \langle-0.81,-0.27)$,
central=cos $\theta \in \langle-0.27,0.27)$ 
and forward=cos $\theta \in \langle 0.27,0.81 \rangle$.
As a reference parameter in our analysis, we will use
the ratio of backward to central emission 
\begin{equation}
bc=\frac{N(backward)}{N(central)}.
\end{equation} 
The choice of the regions was influenced by the fact 
that the experimental efficiency 
of detecting both the heaviest residue and the $\alpha$-particle
in the same detector is small. 
This can be, in fact, seen from the angular
distributions shown in the lower panel of Fig. \ref{figure2},
where we observe the drop of events in the region of $|$cos$\theta | >$0.8.
Therefore, $\alpha$-particles in the angle range 
$|$cos$\theta | >$0.8 were excluded from further analysis.


\section{Theoretical interpretation}


The sequential particle evaporation from hot nuclei is a quite reliable 
approach for the description of experimental data at excitation energies 
E*\loo3-5 MeV/nucleon. In peripheral nucleus--nucleus collisions the 
evaporation process may be very fast and proceed when the PLS and the TLS 
(target-like source) are not far away from each other. 
Therefore, the evaporation of 
particles from both excited sources happens in the common Coulomb field. 
To examine the main features of Coulomb proximity decay, we have constructed 
a model in which the PLS with mass number $A$ and charge $Z$, 
characterized by a spin 
and excitation ($E^{*}$), moves away from TLS with velocity $V$. At a given 
separation distance the de-excitation of PLS via sequential binary decay 
of light particles ($n$, $Z\leq$2) and heavy clusters (up to $^{18}$O) is 
calculated using a Weisskopf approach \cite{Weisskopf}. 
In this model \cite{Botvina87}, the decay width 
for the emission of a particle $j$ 
in the excited state $i$ from a nucleus with ($A,Z$) is given by: 
\be \label{eq:eva}
\Gamma_{j}^{i}=\int_{0}^{E_{CN}^{*}-B_{j}-\epsilon_{j}^{(i)}}
\frac{\mu_{j}g_{j}^{(i)}}{\pi^{2}\hbar^{2}}\sigma_{j}(E)
\frac{\rho_{f}(E_{CN}^{*}-B_{j}-\epsilon_{j}^{(i)}-E)}{\rho_{CN}(E_{CN}^{*})}EdE
\ee
Here $\epsilon_{j}^{(i)}~(i=0,1,\cdots,n)$ are the energies of the ground and 
all particle-stable excited states of the fragment $j$, 
$g_{j}^{(i)}=(2s_{j}^{(i)}+1)$   is  the 
spin degeneracy factor of the $i$-th excited state, 
$\mu_{j}$ and $B_{j}$ are the corresponding reduced mass and separation energy, 
$E_{CN}^{*}$ is the excitation energy of the initial compound nucleus 
(i.e. PLS), and $E$ is the kinetic energy of an emitted particle. 
The level densities of the initial compound $(A,Z)$ and the final $(A_{f},Z_{f})$
residual nucleus, $\rho_{CN}$ and $\rho_{f}$, are calculated using the 
Fermi-gas formula $\rho (E)\propto exp\left(2\sqrt{aE}\right)$ with the 
level density parameter $a \approx 0.15A$MeV$^{-1}$ corresponding to the 
nuclei with $A \approx 20-30$. 
In the present work, we have parametrized the inverse cross section as 
$\sigma_{j}(E)=\pi R_{fj}^{2}(1-U_{c}/E)$, 
where $U_{c}$ is the Coulomb barrier for fragment emission and 
$R_{fj}=R_{f}+R_{j}$, $R_{f}=r_{0}A_{f}^{1/3}$, 
$R_{j}=r_{0}A_{j}^{1/3}$, with $r_{0}=1.5$fm. 

We assume that the electromagnetic interaction between TLS 
and PLS influences each evaporation act. 
The transformation of the Coulomb energy into nuclear one 
(and vice versa) can be described within different approaches. 
Previously, in multifragmentation reactions, this problem was 
resolved by taking into account the whole coordinate phase space 
of the produced fragments \cite{Botvina99}. The phase space analysis is 
model-independent and it can also be applied for our evaporation case. 
However, for better physics understanding  of the phenomenon, it is useful 
to consider a particular model process. Therefore, we suggest that the 
TLS Coulomb field leads to a shift of the proton distribution with respect to 
to the neutron one in PLS. This means that the PLS charge center can 
be shifted from the PLS center-of-mass. A similar process of a 
dipole charge polarization is well known in electromagnetic interactions 
with nuclei as a Goldhaber-Teller and 
Steinwedel-Jensen giant resonances \cite{Goldhaber48,Steinwedel50}. 
Since the evaporation rate is very fast at high excitation energies, 
the charge distribution of the evaporated particle and the residue 
should correspond to this 'dipole' shift of the PLS charge distribution. 
We consider different positions of the charge centers of the evaporated 
daughter's $\alpha$ and the residual by assuming them to be touching charged 
spheres, which are placed within a sphere of radius $R_{j}+R_{f}$ with 
the center located at the center-of-mass of the decaying (PLS) nucleus. 
These coordinate positions are simulated with the Monte-Carlo method. 
In this case the shifts of the PLS charge center ($\sim$1 fm) match 
rather well to the displacements of the 
neutron and proton distributions expected 
in giant resonances \cite{Goldhaber48}. 
The corresponding shift of the neutron distribution of the residue may 
preserve the PLS center-of-mass conservation during the decay. This is 
consistent with the fact that the PLS residue remains excited and it can 
also be dipole-like deformed in the external Coulomb field. 
However, we note that the PLS center-of-mass conservation is not 
an obligatory requirement, contrary to the case of a two-body evaporation 
of an isolated nucleus, because of an energy transfer from the 
external Coulomb to the nuclear deformation of PLS during the decay. 
While the center-of-mass of the whole three-body system (TLS, the residue 
and the daughter fragments) must be conserved during the evaporation, 
as it is in our model. 
In this paper we are limited to a qualitative description of this 
new kind of emission, and we do not develop the model further 
(e.g., there should be a connection between the magnitude of the dipole
polarization and the symmetry energy of nuclei at high excitation energies, which
can be addressed in these processes).
Also, the scope of this paper does not include other consequences of the 
dipole polarization, like a possible neutron enhancement of particles 
emitted toward TLS. 

Within this approach the decay configuration is impacted by the 
change in the Coulomb energy of the three-body system: 
\be 
U_{c}=( \frac{Z_{f}Z_{j}}{R_{fj}}+\frac{Z_{TLS}Z_{j}}{R_{TLS-j}}+
\frac{Z_{TLS}Z_{f}}{R_{TLS-f}}-\frac{Z_{TLS}Z}{R_{TLS-CN}} )e^2, 
\ee
where $Z_{TLS}$ is the charge of TLS, $R_{TLS-CN}$ is an initial 
distance from the TLS to the compound nucleus, and $R_{TLS-j}$ 
($R_{TLS-f}$) is a distance from the TLS to the charge center of 
the emitted particle $j$ (the residue $f$), which takes also into account 
the TLS fluctuations guaranting the center-of-mass 
conservation of the three-body system, respectively. The probability 
of fragment emission can be found by averaging eq.~(\ref{eq:eva}) over 
all coordinates of the charge centers of 
fragments. Since the Coulomb energy is lower when fragment $j$ is 
emitted in the direction of the TLS, the resulting fragment emission 
is anisotropic in the PLS frame. The angular momentum of the PLS 
was included using a standard approach \cite{Botvina01,Charity88}. 

The width for evaporation of particle $j$ is 
$\Gamma_{j}=\sum \Gamma_{j}^{i}$, and the full width is 
determined by summing up 
all emission channels: $\Gamma=\sum \Gamma_{j}$. By definition 
we take the mean time for an emission step as 
$\tau= \hbar / \Gamma$. The PLS is assumed to have a lifetime, $t$, with 
a distribution $exp(-t/\tau)$. Until the decay occurs, the PLS, TLS, 
and all charged particles propagate along Coulomb trajectories. 
Successive binary emissions, conserving energy and momentum, are calculated 
until the excitation energy is below the particle emission threshold. 
In the case of this reaction, we do not take into account the Coulomb 
influence of the previously emitted particles on probabilities of 
subsequent emissions, because the large charge of the TLS dominates. 

\section{Discussion of the results}

\begin{figure}

\resizebox{1.0\textwidth}{!}{
\includegraphics{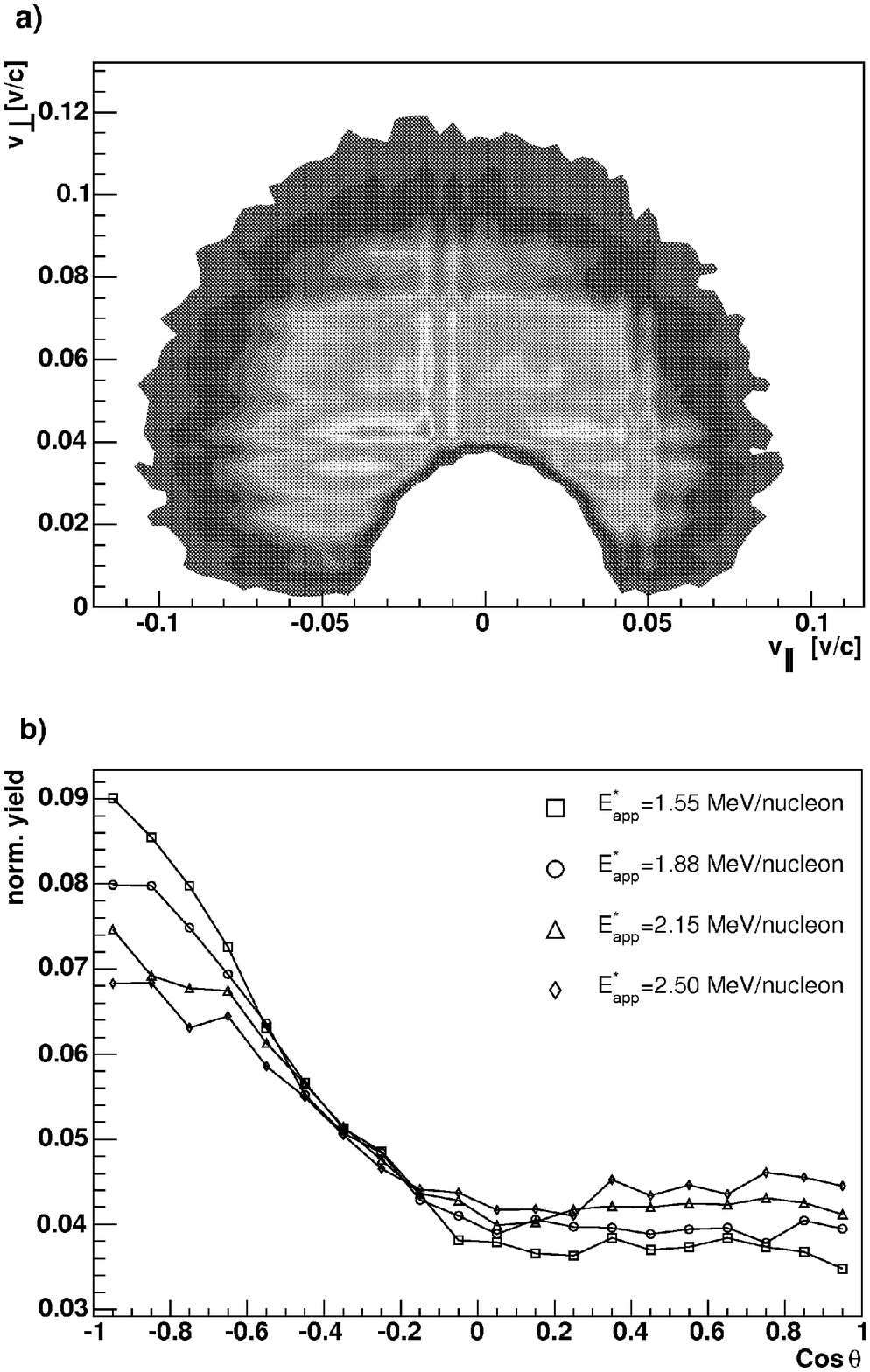}
}

\caption{Velocity scatter plot of $\alpha$-particles (a) and angular distribution
of $\alpha$-particles (b) obtained from calculations of proximity decay of $^{29}$Si
at 20 fm and apparent excitation energies $E^{*}_{app}$=1.55, 1.88, 2.15 and 2.50 MeV.  }
\label{figure3}

\end{figure}

After separation of the PLS and the TLS, the emission rates for particles and 
their angular distributions depend on excitation energies and distances 
between them. Fig. \ref{figure3} shows the calculated scatter plots and 
angular distributions of emitted $\alpha$-particles for different 
excitation energies at a fixed distance (20 fm). 
The effect becomes more pronounced at lower energies since the Coulomb 
energy surplus becomes more essential. However, according to the 
calculations, at low excitation 
energy the decay time is large, so the distance can not be small. In this 
respect, we have a well defined problem: to find the right distances 
at which the decay occurs
in order to describe the observed angular anisotropy. 
Since we know the relative PLS-TLS velocity, one can easily connect them 
with the decay time. The finite size of the TLS and the fluctuations 
of the coordinates of the emitted particle and the residue as a result of the 
evaporation were taken into account. However, the corresponding 
corrections are not 
essential at large distances and do not influence the obtained trends. 

\begin{figure}

\resizebox{1.0\textwidth}{!}{
\includegraphics{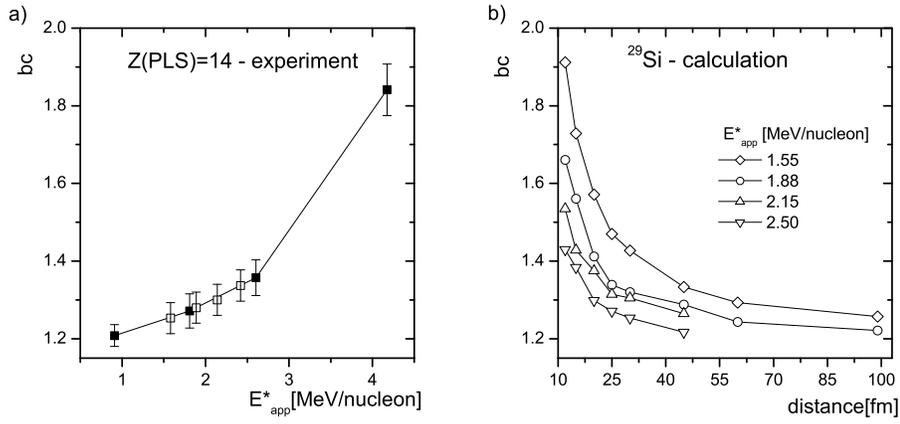}
}

\caption{Anisotropy parameter $bc$ as a function of the apparent excitation energy $E^{*}_{app}$
obtained from experimental data - (a), and calculated values of anisotropy parameter $bc$
as a function of the distance for different $E^{*}_{app}$ - (b). 
Empty squares in (a) show estimated values of $bc$ parameter for $E^{*}_{app}$ involved in the calculation
shown in (b).
}
\label{figure4}
\end{figure}

\begin{figure}

\resizebox{1.0\textwidth}{!}{
\includegraphics{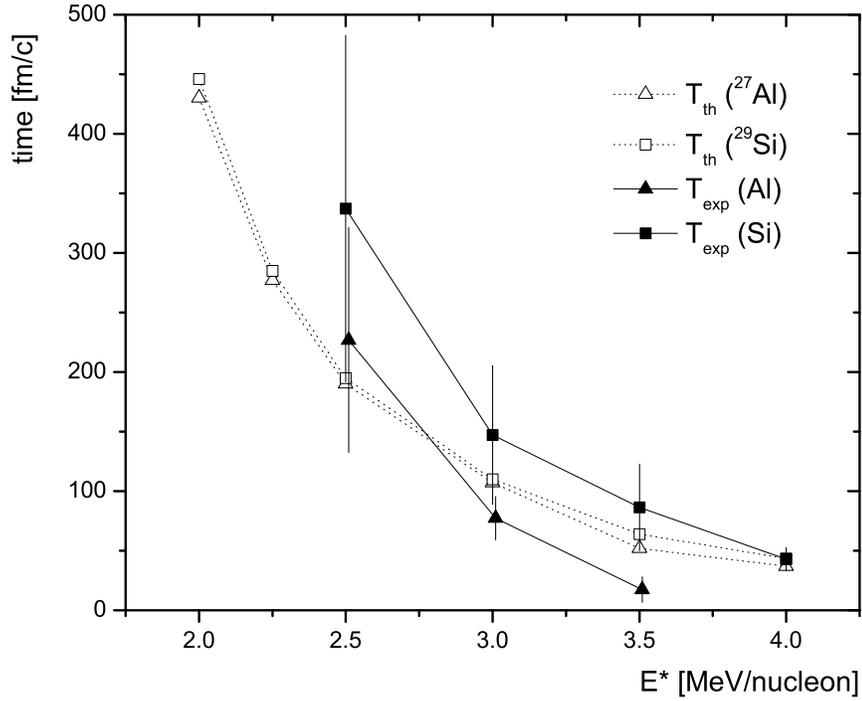}
}

\caption{Time of the first emission $T_{exp}$
obtained by comparison of
angular anisotropy parameter $bc$ observed in experiment (full symbols). 
Time of the emission
$T_{th}$ calculated
from total decay width is shown for comparison (empty symbols). 
Square and triangle symbols correspond to results for Z(PLS)=13 (Si) and Z(PLS)=13 (Al), respectively.}
\label{figure5}
\end{figure}

Fig.~\ref{figure4}a shows the experimental angular anisotropy parameter 
$bc$, observed for different apparent excitation energy bins. 
Fig.~\ref{figure4}b demonstrates how the calculated ratio $bc$ changes with 
the distance at which the first emission (after separation of PLS and TLS) 
happens for four initial excitation 
energies of $^{29}$Si ($E*=$2.5, 3, 3.5 and 4 MeV/nucleon). 
After recalculation of the apparent excitation energy 
according to (\ref{for1}) (the values of apparent excitation energies
in calculation are shown in Fig.~\ref{figure4}b), one can directly
compare the asymmetric parameter values $bc$ and extract the distances
at which the first emission occurs for the corresponding apparent 
excitation energy. 
The time of the emission is then determined
from the separation distances of PLS-TLS divided by their relative 
velocity. 
The time of the first emission is shown in Fig.~\ref{figure5} for Si and Al. 
For illustration, 
we show the evaporation times calculated according to the widths found 
in the model for $^{29}$Si and $^{27}$Al. 
A rapid lifetime drop with the excitation is a general behavior 
of evaporation lifetimes (see also, e.g., \cite{Charity00}), and it 
reflects a gradual transition to simultaneous multifragment decay, which 
has been discussed in the literature \cite{ISIS}. One can see a qualitative 
agreement for the extracted and the calculated times. 
Improvements in model parameters and selection of an isotopic distribution
of sources should result in even better agreement, however the primary goal 
of the present study is to demonstrate the possibility of this method. 

We emphasize, that in our analysis 
we exclude $\alpha$-particles with $v>|0.1 c|$, 
in order to guaranty a selection of an evaporation-like 
process. The energy spectra of these particles can be explained by a normal 
evaporation emission (see also \cite{Hudan03}). While one 
may speculate that the asymmetry is induced by the previous dynamical 
stage there are currently no dynamical models which are able to 
describe this process. The dynamical emission is usually associated with 
particle production at midrapidity velocities, which are excluded from 
our analysis. We expect that the time for a dynamical emission is 
rather short $\loo$50 fm/c, while the present method is aimed at the decay 
times $\approx$50-500 fm/c, where an equilibration is quite probable. 
Additionally, equilibrium is to be expected because of the relatively 
low energy released during emission of one particle 
(i.e. just slightly above the threshold energy for 
$\alpha$ emission), in comparison with a large energy transferred during 
the collision process. Therefore, it is a reasonable explanation that 
this phenomenon can occur as a joint effect of nuclear and electromagnetic 
forces during the fast decay of PLS in the proximity of TLS.

We note that the Coulomb proximity effect
of evaporative alphas in the direction between the two larger fragments was 
already reported
for fission events \cite{Brucker87}. However, in this previous work it was
not possible to explain fully the magnitude of this effect.
In our new interpretation an enhancement of
the 'midrapidity' evaporation is connected
with a dipole-like charge deformation inside the
nucleus, which can lead to the creation of  a new minimum in
the potential energy at the moment of evaporation
in the presence of the third charged body.
In this way, the external Coulomb energy is
converted into internal energy of the source.

\section{Conclusions}
We have considered the evaporation-like decay of highly excited 
projectile nuclei produced in peripheral nucleus-nucleus collisions. 
The time for production of these nuclei can be estimated as $\leq$50 
fm/c (the time of separation of the projectile and target 
nuclei). Afterwards, the nuclei undergo rapid de-excitation. We 
demonstrated that this decay has a large angular asymmetry, which is 
not expected for the standard evaporation process. As was pointed out, 
the physical condition at which this evaporation happens is different 
from the evaporation from an isolated compound nucleus. Since the decay 
is very fast the target is still in the vicinity of the decaying 
projectile nucleus, and their Coulomb interaction influences the 
decay. This effect can explain the observed asymmetry, as was shown 
by comparison with our model. One can connect the observed asymmetry 
with the distance between the target and the projectile and 
consequently with the decay time. In this respect, the "proximity 
decay" may be considered as a "clock" sensitive to the 
times of 50-500 fm/c, during which this proximity exists. 
To our knowledge, these are the smallest 
decay times which can be identified in a sequential framework. 
At higher excitation 
energies, we expect a simultaneous multifragment decay into many 
small fragments. 

\section{Ackowledgements}
The authors wish to thank the staff of the Texas A\&M Cyclotron facility
for the excellent beam quality. This work was supported in part by the Robert
A. Welch Foundation through grant No. A-1266, and the Department of Energy
through grant No. DE-FG03-93ER40773.
We thank R.~De~Souza, L.~Sobotka, R.~Charity and S.~Hudan for 
stimulating discussions.


\begin{thebibliography}{00}

\bibitem{Bertulani88} 
C.A.Bertulani and G.Baur, {\em Phys. Rep.}, {\bf 163}, 299 (1988).

\bibitem{Pshenichnov98}
I.A.~Pshenichnov et al., {\em Phys. Rev.} {\bf C57}, 1920 (1998)

\bibitem{Botvina99} A.S.Botvina, M.Bruno, M.D'Agostino and D.H.E.Gross, 
 {\em Phys. Rev.} {\bf C59}, 3444 (1999).

\bibitem{Botvina01} A.S.~Botvina and I.N.~Mishustin, Phys. Rev. {\bf C63}, 
061601(R) (2001). 

\bibitem{Weisskopf}
 V. Weisskopf, Phys. Rev., {\bf 52}, 295 (1937)

\bibitem{Ericson} T.~Ericson, Adv. in Phys., {\bf 9}, 425 (1960). 

\bibitem{Hudan03}
S.~Hudan et al., {\em nucl-ex/0308031}, (2003)

\bibitem{Laforest99} A.~Laforest et al., {\em Phys. Rev.} {\bf C59}, 
2567 (1999). 

\bibitem{Botvina87}
 A.S. Botvina et al., Nucl. Phys. {\bf A475}, 663 (1987).

\bibitem{Goldhaber48} M.~Goldhaber and E.~Teller, 
Phys. Rev. {\bf 74}, 1046 (1948). 

\bibitem{Steinwedel50} H.~Steinwedel and H.~Jensen,
Z. Naturforsch. {\bf 5A}, 413 (1950).

\bibitem{Charity88} R.J. Charity et al., Nucl. Phys. {\bf A483}, 371 (1988).

\bibitem{Charity00} R.J. Charity et al., Phys. Rev. {\bf C61}, 054614 (2000).

\bibitem{ISIS} L.~Beaulieu et al., Phys. Rev. Lett. {\bf 84}, 5971 (2000). 

\bibitem{Brucker87} A.~Brucker et al., Phys. Lett. B, {\bf 186}, 20 (1987).

\end{thebibliography}

\newpage


\pagebreak


\pagebreak


\pagebreak


\end{document}